\documentclass[twocolumn,prd,footinbib,aps,prl,floats,floatfix,amsmath,amssymb,longbibliography,secnumarab
ic]{revtex4-1} %

\usepackage[final]{graphicx}
\usepackage{hyperref}
\usepackage{amsmath}
\usepackage{bbm}
\usepackage{amsfonts}
\usepackage{amssymb}
\usepackage{latexsym}
\usepackage{graphicx}
\usepackage[english]{babel}
\usepackage{multirow}
\usepackage{float}
\usepackage{url}
\usepackage{slashed}
\usepackage{xcolor} 
\usepackage[utf8]{inputenc}
\usepackage{verbatim}
\usepackage{stackengine}
\usepackage{cancel}
\usepackage{accents}
\usepackage{array}
\usepackage{physics}
\usepackage{bm}

\def\sfrac#1#2{{\textstyle{#1\over #2}}}

\newcommand{\be}{\begin{equation}}
\newcommand{\ee}{\end{equation}}
\newcommand{\ba}{\begin{array}}
\newcommand{\ea}{\end{array}}
\newcommand{\bea}{\begin{eqnarray}}
\newcommand{\eea}{\end{eqnarray}}

\usepackage{slashed}

\definecolor{darkgreen}{rgb}{0,0.5,0}

\begin{document}

\title{Axion strings from string axions}
\author{James M.\ Cline}
\email{jcline@physics.mcgill.ca}
\thanks{ORCID: \href{https://orcid.org/0000-0001-7437-4193}{0000-0001-7437-4193}}
\affiliation{McGill University Department of Physics \& Trottier Space Institute, 3600 Rue University, Montr\'eal, QC, H3A 2T8, Canada}
\affiliation{Niels Bohr International Academy,The Niels Bohr Institute,
Blegdamsvej 17, DK-2100 Copenhagen Ø, Denmark}
\author{Christos Litos}
\email{c.litos@ufl.edu}
\author{Wei Xue}
\email{weixue@ufl.edu}
\affiliation{Institute for Fundamental Theory, Physics Department, University of Florida,
Gainesville, FL 32611, USA}

\begin{abstract}
A favored scenario for axions to be dark matter is for them to form a cosmic string network that subsequently decays, allowing for a tight link between the axion mass and relic abundance.  We discuss an example in which the axion is
protected from quantum gravity effects that would spoil its ability to solve the strong CP problem: namely a string theoretic axion arising from gauge symmetry in warped extra dimensions.
Axion strings arise following the first-order Randall-Sundrum compactification phase transition, forming at the junctions of three bubbles during percolation.
Their tensions are at the low scale associated with the warp factor, and are parametrically smaller than the usual field-theory axion strings, relative to the scale of their decay constant.
Simulations of string network formation by this mechanism must be carried out to see whether the axion mass-relic density relation depends on the new parameters in the theory.

\end{abstract}
\maketitle
{\bf  1.\,Introduction}.
The axion solution to the strong CP problem \cite{Peccei:1977hh,Kim:2008hd,Weinberg:1977ma,Wilczek:1977pj,Peccei:2006as} provides a highly motivated dark matter candidate \cite{Abbott:1982af,Preskill:1982cy}, potentially resolving two of the major mysteries of particle physics through one hypothesis. Conventional axion models involve a complex scalar field $\Phi = \varphi\, e^{ia/f_a}$, the Peccei-Quinn (PQ) field, whose angular component $a$ is the axion.  It is initially a massless Goldstone boson when the global PQ symmetry is spontaneously broken by the vacuum expectation value $\langle\varphi\rangle = f_a$ (known as the axion decay constant).  This occurs at high temperatures in the early Universe.  At lower temperatures, below the QCD phase transition, the axion gets a small mass because the PQ symmetry is anomalous, and the axion couples to the gluon field strength via 
$(a/f_a) G\tilde G$. 

In terms of dark matter predictions, a favored scenario is for the PQ phase transition temperature to be below the reheating temperature after inflation.
In this case, a cosmic string network forms, in which $a/f_a$ winds around each string by a multiple of $2\pi$.  These strings eventually decay into axion particles, and produce a relic density that depends on the single parameter
$m_a$, the axion mass.  Then in principle, experimentalists searching for relic axions could concentrate their efforts on the value of $m_a$ that leads to the observed dark matter abundance.  In practice, the string network computations needed to make a precise prediction of $m_a$ are numerically quite demanding, and there is not yet a consensus on the range of likely values, although much progress has been made. 

Quantum gravity poses a generic problem for field theory axions: it is believed to break global symmetries, including the PQ symmetry, which would spoil the axion solution to the strong CP problem unless one artificially forbids PQ-breaking operators up to some very high dimension \cite{Barr:1992qq,Holman:1992us,Kamionkowski:1992mf}.  On the other hand, string theories naturally predict light axions that can be protected from such corrections, for example by having the global symmetry descend from an underlying gauge symmetry. In this case, the symmetry-breaking contributions from quantum gravity are expected to be exponentially suppressed \cite{Craig:2024dnl}.
Recently Ref.\ \cite{Benabou:2023npn} studied string theoretic axions, and 
concluded that their cosmic strings have tensions near the Planck scale.  They therefore correspond to conventional theories with PQ-breaking above the scale of inflation, which never form string networks.  These models rely upon the misalignment mechanism \cite{Kibble:1976sj,Zurek:1985qw} for getting the axion relic density, which provides no direct link between $m_a$ and the dark matter abundance.

In this work, we discuss a simple counterexample to the above statement, starting from a 5-dimensional Randall-Sundrum \cite{Randall:1999ee} setup with a warped extra dimension and a U(1) gauge field living in the bulk.  The axion is the zero mode of $A_5$, the 5th component of the gauge field; hence it is protected from quantum gravity corrections by the gauge symmetry \cite{Choi:2003wr}.  Since the starting point does not involve any
explicit PQ scalar field $\Phi$, it is not {\it a priori} obvious what plays the role of the radial component $\varphi$ that determines the cosmic string profile.
The answer turns out to be that $\varphi$ is the radion, the light degree of freedom that corresponds to fluctuations in the size of the extra dimension.
Deep in the core of the string, the radion vanishes, corresponding to a phase in which the field of the low-energy description have dissolved into more elementary constituents, analogous to the quark-hadron phase transition of QCD.

An important difference relative to conventional cosmic strings is the formation mechanism, which is usually a second order phase transition from a high-temperature $T^2|\phi|^2$ potential to a low-temperature Higgs-like potential 
$\lambda(|\phi|^2-v^2)^2$.  In the 5D setup, the extra dimension is effectively decompactified  at high-T, and becomes compact during a strongly supercooled first order transition where bubbles of the broken phase are nucleated.  Each bubble contains a region where $\langle a\rangle/f_a$ takes a random value in the interval $[0,\,2\pi]$.  As the phase transition completes, bubbles  coalesce, and at a junction of three bubbles in which $a$ approximately winds around the circle enclosing the junction, a cosmic axion string can form.  The dynamics of such string formation is very different from those in the second order transition, and have not yet been studied in detail.  We do not attempt such a simulation here, but rather lay the groundwork for future studies.

\medskip
{\bf 2.\,Framework.} The geometry of the Randall-Sundrum model can be described by a warped 5D anti-deSitter
metric
\be
    ds^2 = \left({\varphi\over F}\right)^{2u}\eta_{\mu\nu}dx^\mu dx^\nu + {\ln^2(\varphi/F)\over k^2}\,du^2\,,
\label{RSmetric}\ee
having an ultraviolet (UV) brane at $u=0$ and an infrared (IR) brane at $u = 1$.
The field $\varphi$ is the radion, which describes dynamical fluctuations of the size of the extra dimension, 
and $F =\sqrt{24 M_5^3/k} \cong \sqrt{24} M_p$, with
$M_5$ being the 5D gravity scale, $k$ the AdS curvature scale, and $M_p$ the 4D 
Planck scale.  The effective 4D description of gravity is general relativity 
plus small higher derivative corrections if $M_5/k$ is sufficiently large.
The radion is naturally stabilized by a potential $V(\varphi)$ at a vacuum expectation value $\langle\varphi\rangle = v \ll k$, which achieves a large hierarchy of scales between the IR and UV branes without the need for fine tuning.
Originally $v$ was taken to be the TeV scale to explain the standard model weak scale, but here we are interested intermediate scales such as $v\sim 10^{10}$\,GeV, corresponding to the scale of axion couplings (even though there is no explicit global PQ symmetry in the theory).

The radion plays an important role in the RS phase transition at high temperatures, in which the theory is effectively decompactified.  It starts near $\varphi=0$,
and at the critical temperature it makes a first order transition to its vacuum expectation value $v$.  
Above the transition temperature $T_c$, the radion is no longer a good effective 
description; rather there is a strongly coupled conformal SU(N) theory, and it becomes confining at $T_c$. 

Generically, the transition can be so supercooled that it never completes \cite{Creminelli:2001th}, especially in regions of parameter space where the 4D effective description is under control.  However it is possible to overcome this obstacle through careful design of the radion potential \cite{Agashe:2020lfz}, which comes from the Goldberger-Wise mechanism \cite{Goldberger:1999uk}, {\it i.e.,} integrating out a bulk scalar field whose couplings to $\varphi$ induce its potential:
\be
    V(\varphi) = \lambda\left[\varphi^4 f(\varphi/v) -v^4 f(1)\right]\,,
\ee
where $x^4 f(x)$ is a function with a minimum at $x=1$, whose form depends upon details of the GW bulk scalar field Lagrangian.  We consider two typical
examples, denoted by CMS \cite{Chacko:2013dra} and GW \cite{Goldberger:1999uk},
respectively:
\be
    f(x) = {x^\epsilon\over (1+\epsilon/4)} -1,\quad
        f(x) = (1-x^\epsilon)^2\,.
\ee
The small exponent $\epsilon$ relates the large hierarchy of scales between $M_p$ and $f_a$ to a moderately small ratio of scales in the GW field potential being raised to a large power $1/\epsilon$; we take $\epsilon = 0.1$ for definiteness in numerical examples.  The coupling $\lambda$ is expected to be of order
$(k/M_5)^6$, which must be small for the effective theory to be trustworthy.
The radion's  squared mass is given by $m_\varphi^2 = V''(v) = 4\lambda\epsilon v^2$, $2\lambda\epsilon^2 v^2$ for the two respective potentials.  Its kinetic term is simply $\sfrac12\partial_\mu\varphi\,\partial^\mu\varphi$, apart from small
corrections of order $(\varphi/F)^2$ that we neglect.

\begin{figure}[t]
\centerline{
\includegraphics[width=0.48\textwidth]{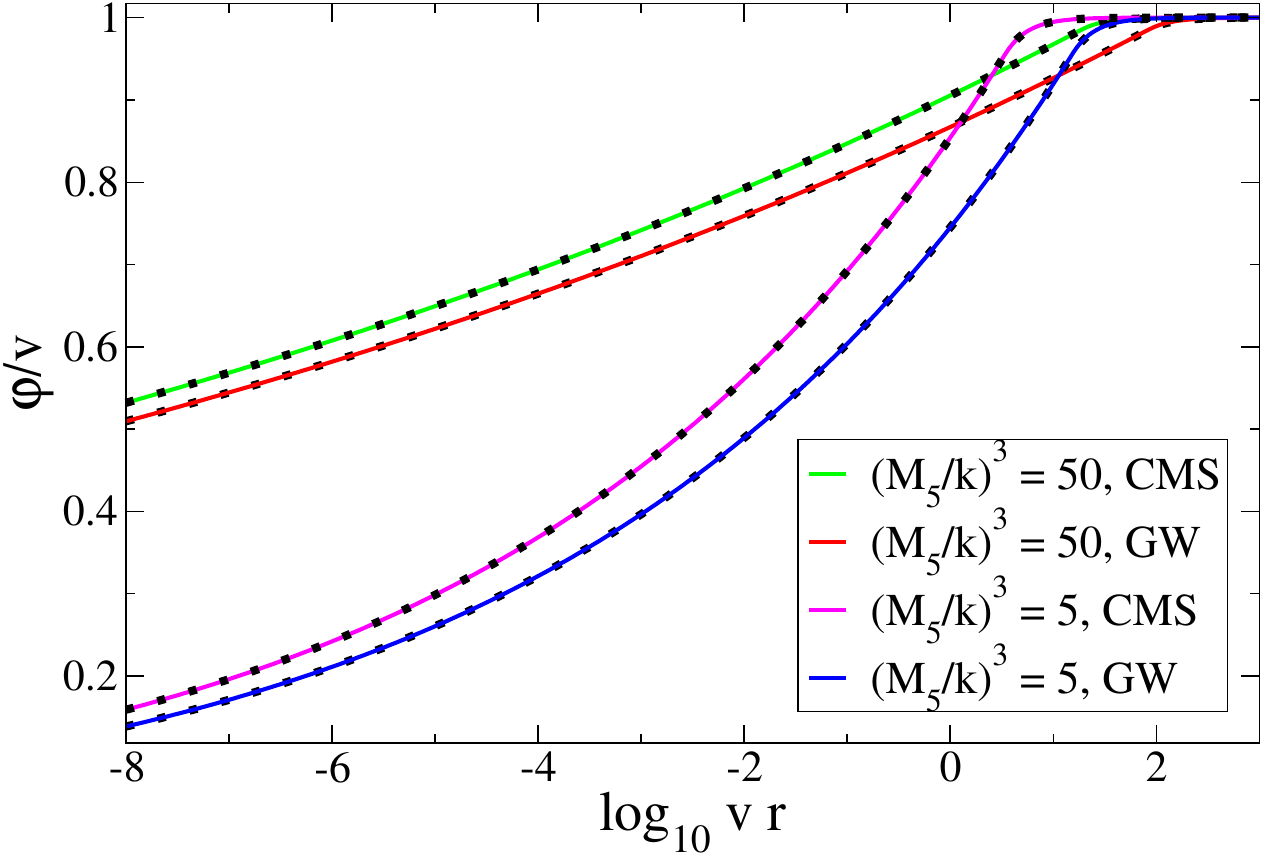}}
    \caption{Left: example cosmic string profiles $\varphi(r)$ for two choices of $(M_5/k)^3$ and the radion potential (CMS \cite{Chacko:2013dra} and GW \cite{Goldberger:1999uk}), in units of the radion VEV $v$. Solid curves are numerical solutions, dotted are the analytic approximation (\ref{asol}).  Other parameters are taken to be
    $g_5^2 k = 1$, $\lambda = 1/(M_5/k)^6$, $\epsilon = 0.1$.}
    \label{soln-plots}
\end{figure}

To incorporate an axion in the theory, we follow Refs.\ \cite{Choi:2003wr,Flacke:2006ad} by 
introducing a U(1) gauge field $A_M$ in the bulk, with $A_5$ playing the role of the massless axion.
We impose the Dirichlet boundary conditions for $A_\mu$ on the UV and IR branes, so that the massless $A_\mu$ is eliminated. The massive components of $A_\mu$ are integrated out  by solving the equation of motion. This gives the relation $F_{\mu 5} =  \ln(F/\varphi) (F/\varphi)^{2(u-1)} \partial_\mu \vartheta(x)$, where $\vartheta(x) \equiv 2  \int_0^1du\ A_5$ (see Appendix A for details). In the string solution with a winding number of 1, $\vartheta(x)$ also represents the azimuthal angle in the plane transverse to the string.
From this, we derive the effective Lagrangian for $\vartheta(x)$
\footnote{It simplifies the calculation to choose an axial gauge $A_5 =$ constant, where the constant is quantized by the requirement of the Wilson
loop being a multiple of $2\pi$.},
\be 
    {\cal L}_a = {k\over 2 g_5^2 F^2}\, \varphi^2\, \partial_\mu \vartheta  \, \partial^\mu\! \vartheta  \equiv  \sfrac12 \left(\varphi\over v\right)^2 \partial_\mu a  \, \partial^\mu\!{a}\,,
\ee 
where the 5D gauge coupling $g_5$ has dimensions of (mass)$^{-1/2}$. 
The decay constant $f_a$ is determined by the periodicity of $a$ in field space, which depends on that of $A_5$. It follows from the Wilson loop 
$\exp(2i\int_0^1 du A_5)$ which is equal to unity for $A_5 = \pi n$, 
implying that 
\be
    f_a = {\sqrt{k} \, v \over g_5 \, F}\,,
    \label{faeq}
\ee
in agreement with Ref.\ \cite{Benabou:2023npn}: $f_a$ is at the warped scale of the IR brane.

As will be seen, this result for $f_a$ is the one that determines the cosmic string tension, but it does not automatically determine the coupling strength of axions to matter. 
Ref.\ \cite{Flacke:2006ad} shows that it depends on whether
QCD resides on the branes or in the bulk.  Only in the latter case, the axion  coupling to gluons is of order $1/f_a$
(\ref{faeq}); otherwise the coupling is suppressed by the Planck scale, spoiling the axion solution to the strong CP problem.  Having a heavy quark with quantized $g_5$ charge in the bulk and integrating it out generates a Chern-Simons interaction between
$A_M$ and the QCD gluons, resulting in the axion-gluon coupling of order {$g^2/( 32\pi^2 f_a ) a G\tilde G$.

\medskip
{\bf 3.\,String Solutions.}
For configurations with axion winding by $2\pi$, the energy density of the string is
\be
   {\cal E} =  \sfrac12\left( \varphi'(r)^2 + b^2 {\varphi^2\over r^2}\right) + V(\varphi)\,,
    \label{eden}
\ee
leading to the equation of motion
\be
    \varphi'' + {\varphi'\over r} = b^2 {\varphi\over r^2} + V'(\varphi)\,,
\ee
where $b^2 = (f_a/v)^2 = [24\, (g_5^2 k) (M_5/k)^3 ]^{-1}$, which is expected to be
small, $b^2\ll 1$.  The exact solution is very well approximated by joining  its 
small- and large-$r$ limits at a matching radius $r_m = \sqrt{2b+b^2}/m_\varphi$:
\be
    \varphi(r) \cong \left\{\begin{array}{ll} v(r/r_m)^b/(1+b/2),& r < r_m\\
    v[1 - b^2/(m_\varphi r)^2],& r > r_m\end{array}\right.\,.
    \label{asol}
\ee
    \label{F}
As examples, we show solutions for $(M_5/k)^3 = 5,\, 50$, $g_5^2k=1$, $\lambda = (M_5/k)^{-6}$, $\epsilon = 0.1$ in Fig.\ \ref{soln-plots}.  The profiles for the 
GW and CMS radion potentials differ only because their respective radion masses
are not the same.

The tension of the cosmic string is found by integrating the energy density
(\ref{eden}), $T = 2\pi\int_0^R dr\, r\, {\cal E}$.
$R$ is the usual infrared cutoff on global cosmic strings, because of the logarithmic growth of the winding term; this is typically cut off by the separation between neighboring strings.   The tension can be divided into 
that coming from the core region, $r < r_m$, and that from the exterior region.
The exterior contribution is
\be
    T_{>} = \pi (bv)^2\left({b\over 4(1+b/2)} + \ln {R\over r_m}\right)\,.
\ee
The first term comes from the radion potential, by approximating it as 
$V\cong \sfrac12 m_\varphi^2 (v-\varphi)^2$ in the vicinity of its minimum, while the second comes from $b^2 v^2/r^2$.  The core contribution to the tension is
\be
    T_c = {\pi b v^2\over (1 + b/2)^2} + O(b^3 v^2)\,,
\ee
where $O(b^3 v^2)$ is the part coming from the radion potential, that turns out to be independent of $\lambda$ and $\epsilon$ for small $b$, with coefficient $5/8$ for CMS and $5/168$ for GW.

It is interesting to contrast the relation between tension and decay constant in this model with that of a generic field theory PQ string, whose Lagrangian is
$|\partial\Phi|^2 - \lambda(|\Phi|^2 -f_a^2/2)^2$, with $\Phi = (\varphi/\sqrt{2}) e^{i a/f_a}$.  This corresponds to the warped axion model when $b=1$ and $m_\varphi^2 = 2\lambda v^2$.  The cosmic string tension is smaller than its field theory counterpart since $b^2 \ll 1$ and $m_\varphi$ is suppressed by $\epsilon$.

\begin{figure}[t]
\includegraphics[width=0.48\textwidth]{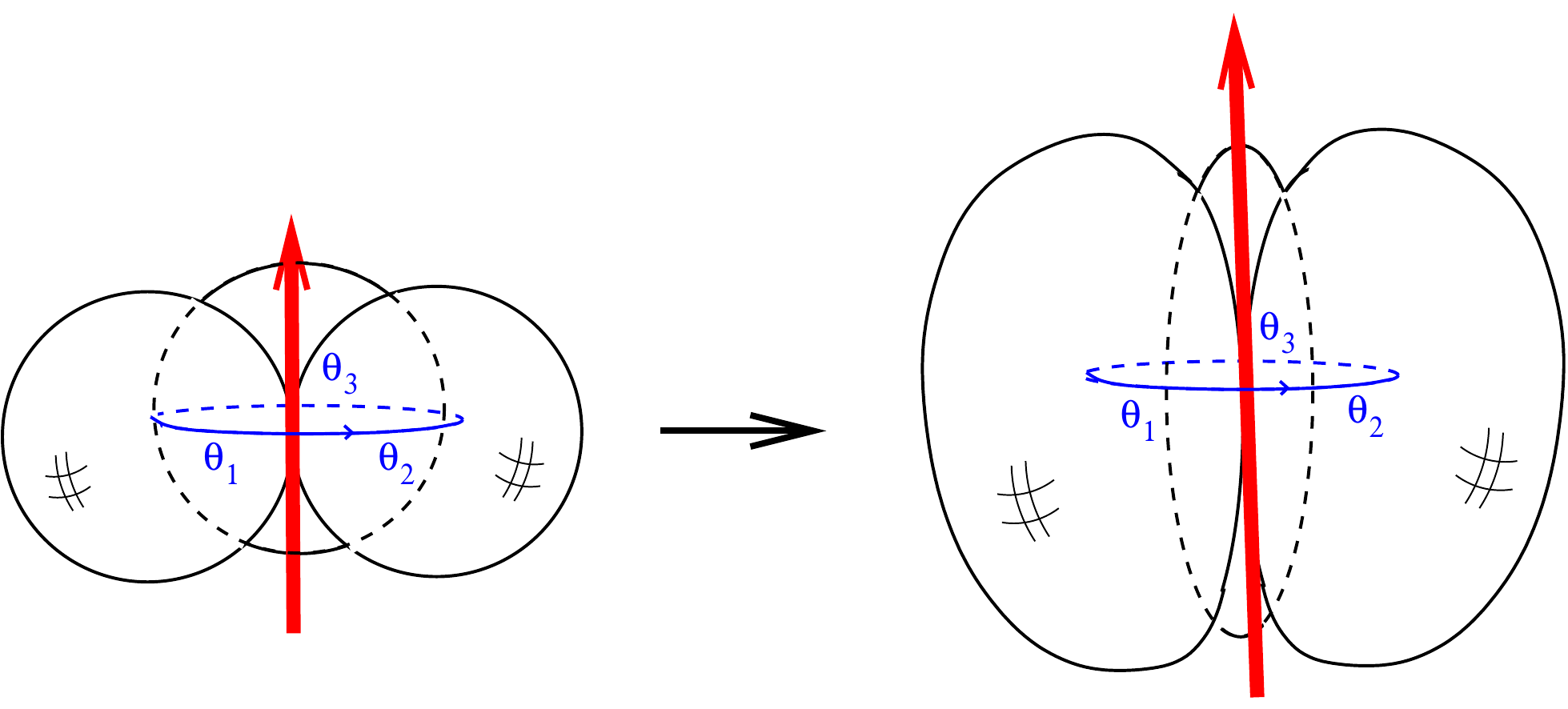}
    \caption{Intersection of three phase transition bubbles, which can initiate cosmic string formation (red arrow) if their axion phases wind around the circle.  The string elongates as the bubbles expand.}
    \label{fig-pt}
\end{figure}

\medskip
{\bf 4.\,String production.}
Unlike the field theoretic PQ phase transition, the RS transition is first order, proceeding by bubble nucleation.  Inside each bubble, the axion field will have an approximately uniform phase $\theta_i = a_i/f_a$.  If three bubbles meet such that $\theta_1 < \theta_2 < \theta_3$, or vice versa, with the angles approximately covering the circle, then the axion is in a winding configuration
and a cosmic string will start to form in the region between the intersections of the bubbles, Fig.\ \ref{fig-pt}.  As the bubbles expand, the interstitial region is
topologically restricted to maintain $\varphi = 0$, and the string grows in length
until it crosses the horizon, or meets another string that it can join with, possibly forming a loop.

This formation mechanism is qualitatively similar to, but different in detail from the Kibble mechanism, which relies upon the tiling of horizon-sized regions, in each of which $\theta$ is approximately constant, rather than the collisions of bubbles.  Generically, we expect the probability of string formation to be lower in the bubble collision picture.  Whereas it is generic for three Hubble patches to be adjoining, since the second order phase transition occurs at the same time everywhere, in the first order transition it is more likely for two bubbles to collide than three.  The phases may have enough time to equilibrate between two bubbles before they meet a third, thereby avoiding winding.  Depending upon how much supercooling occurs, bubble nucleation could be sufficiently rare that three-bubble collisions are suppressed.  Once there is
such a collision, the probability of having a winding configuration is naively $3!/3^3 = 2/9$.

It is possible that the details of the initial production become unimportant once the string network enters a scaling regime.  But even if this occurs, there may still be interesting differences relative to the field theory axion when it comes to the decays of the strings into axion particles.  This motivates numerical simulations of the formation and decay of such networks to supplement their PQ field theory counterparts, to find out whether the relationship between the ultimate axion mass (acquired during the QCD phase transition) and its relic density becomes dependent on new parameters such as $b$ or the degree of supercooling.  Thus far, the production of global strings in a first order phase transition has only been studied in an M.Sc.\ thesis \cite{Vihonen}, but further work in this direction is in progress \footnote{D.\ Weir, private communication}.

\bigskip
{\bf Acknowledgements.}  
We thank Quentin Bonnefoy, Simon Caron-Huot, Kiwoon Choi, Majid Ekhterachian, Tony Gherghetta, Ben Girpaios, Peizhi Du, Noam Levi, Pierre Sikivie, Minho Son and David Weir for valuable discussions. JC thanks the Niels Bohr International Academy for its generous hospitality. JC and WX thank CERN for providing a stimulating working environment that led to this study.
JC is supported by the Natural Science and Engineering Research Council (NSERC) of Canada.
CL and WX are supported in part by the U.S. Department of Energy under grant DE-SC0022148 at the University of Florida.

\appendix 

\section{Kinetic term of 5D gauge field}

In this appendix we integrate out the 5-dimensional U(1) gauge field to obtain the kinetic term for the axion field.   The relevant action is
\begin{equation}
S_{U(1)} = - 2\cdot \frac{1}{4g_5^2} \int d^5x \sqrt{-g}\ g^{MN} g^{PQ} F_{MP} F_{NQ}\,,  
\label{A1}\end{equation}
where the extra factor of two accounts for the $u \leftrightarrow -u$ orbifold identification. In what follows, we assume that $A_\mu$ depends only on the extra-dimensional coordinate $u$ and satisfies Dirichlet boundary conditions on the boundaries, $A_\mu(u=0) = A_\mu (u=1) = 0$. This means that the 4D field is not dynamical and can be integrated out since its kinetic term vanishes, $F_{\mu\nu} = 0$.

The action in (\ref{A1}) thus reduces to  
\begin{equation}
    S_{U(1)} = - \frac{1}{g_5^2} \int d^5x\ \sqrt{-g}\  g^{\mu\nu} g^{55} F_{\mu5} F_{\nu 5}\,.
\label{A2}\end{equation}
Varying with respect to $A_\mu$ and $A_5$ yields the equations of motion (EOMs)
\begin{equation}
    \partial_5\left[\left(\frac{\varphi}{F}\right)^{2u} \frac{\ln(\varphi/F)}{k} F_{\nu 5} \eta^{\nu \mu}\right] = 0\,,
\label{A3}\end{equation}
\begin{equation}
    \partial_\mu\left[\left(\frac{\varphi}{F}\right)^{2u} \frac{\ln(\varphi/F)}{k} F_{\nu 5} \eta^{\nu \mu}\right] = 0\,,
\label{A4}\end{equation}
respectively. 

We first consider Eq.\ (\ref{A3}). The general solution can be written as 
\begin{equation}
    F_{\mu 5} = \left( \frac{\varphi}{F} \right)^{-2u} \frac{\ln(\varphi/F)}{k} f_\mu(x)\,. 
\label{A5}\end{equation}
Integrating this with respect to $u$ and taking the $\varphi \ll F$ limit yields the solution for $f_\mu$:
\begin{equation}
    f_\mu = -2k \left( \frac{\varphi}{F} \right)^2 \partial_\mu \left[ \int_0^1 du\, A_5\right ]\,. 
\label{A6}\end{equation}
The dimensionless axion field $\vartheta$, whose periodicity is $2\pi$, is defined through the gauge-invariant Wilson loop \cite{Arkani-Hamed:2003xts,Benabou:2023npn},
\begin{equation}
    \vartheta(x ) = 2 \int_0^1du\ A_5 \,.
\label{A7}\end{equation}
The periodicity comes from the fact that $e^{i\vartheta}$ is a large 
gauge transformation when $\vartheta$ is a multiple of $2\pi$.
Substituting (\ref{A7},\ref{A6}) into (\ref{A5}) allows us to write $F_{\mu 5}$ as a function of the dimensionless axion field,
\begin{equation}
    F_{\mu 5} =  \ln(F/\varphi) \left( \frac{F}{\varphi} \right)^{2(u-1)} \partial_\mu \vartheta(x)\,. 
\label{A8}\end{equation}

Having obtained an expression for $F_{\mu 5}$ that satisfies (\ref{A3}), all that remains is to check whether there are any further conditions imposed by (\ref{A4}). Substituting into (\ref{A4}) yields: 
\begin{equation}
    \partial_\mu \left[ \left(\frac{\varphi}{F} \right)^2 \partial^\mu \vartheta(x) \right] = 0\,, 
\label{A9}\end{equation}
which is satisfied, provided we identify $\vartheta(x)$ with the azimuthal angle in the cosmic string solution, 
while assuming a radial dependence for the radion $\varphi$.
\bibliography{main}
\bibliographystyle{utphys}

\end{document}